\documentclass[aps,twocolumn,pra,floatfix,superscriptaddress,showpacs,longbilbiography]{revtex4-1}
\usepackage[utf8]{inputenc} 

\usepackage{graphicx}
\usepackage{dcolumn}
\usepackage{bm}
\usepackage{comment}

\usepackage{braket}
\usepackage{amsfonts}
\usepackage{amsmath}
\usepackage{amssymb}

\usepackage{color,soul}
\usepackage{wasysym}
\usepackage{mathrsfs}
\usepackage{times}

\usepackage[dvipsnames,svgnames,table]{xcolor}
\usepackage{hyperref}
\usepackage{fancyhdr}
\usepackage{float}
\usepackage[english]{babel}
\usepackage[caption=false]{subfig}

\hypersetup{
pdfstartview={FitH},% fits the width of the page to the window
colorlinks=true,    % false: boxed links; true: colored links
linkcolor=blue, % color of internal links
citecolor=blue,   % color of links to bibliography
filecolor=blue, % color of file links
urlcolor=blue   % color of external links
}

\newcommand{\bea}{\begin{eqnarray}}
\newcommand{\eea}{\end{eqnarray}}

\newcommand{\quotes}[1]{``#1''}

\begin{document}
\title{Edge states in trimer lattices}

\author{V. M. Martinez Alvarez}%$^*$
\affiliation{Departamento de F\'{\i}sica, Laborat\'orio de F\'{\i}sica Te\'orica e Computacional, Universidade Federal de Pernambuco, Recife 50670-901, Pernambuco, Brazil}
\author{M. D. Coutinho-Filho}%$^*$
\affiliation{Departamento de F\'{\i}sica, Laborat\'orio de F\'{\i}sica Te\'orica e Computacional, Universidade Federal de Pernambuco, Recife 50670-901, Pernambuco, Brazil}

\begin{abstract}
Topological phases of matter have attracted much attention over the years. Motivated by analogy with photonic lattices, here we examine the edge states of a one-dimensional trimer lattice in the phases with and without inversion symmetry protection. In contrast to the  Su-Schrieffer-Heeger model, we show that the edge states in the inversion-symmetry broken phase of the trimer model turn out to be chiral, i.e., instead of appearing in pairs localized at opposite edges they can appear at a \textit{single} edge. Interestingly, these chiral edge states remain robust to large amounts of disorder. In addition, we use the Zak phase to characterize the emergence of degenerate edge states in the inversion-symmetric phase of the trimer model. Furthermore, we capture the essentials of the whole family of trimers through a mapping onto the commensurate off-diagonal Aubry-Andr\'{e}-Harper model, which allow us to establish a direct connection between chiral edge modes in the two models, including the calculation of Chern numbers. We thus suggest that the chiral edge modes of the trimer lattice have a topological origin inherited from this effective mapping. Also, we find a nontrivial connection between the topological phase transition point in the trimer lattice and the one in its associated two-dimensional parent system, in agreement with results in the context of Thouless pumping in photonic lattices.
\end{abstract}

%model an effective bulk-boundary correspondence for the phase with broken inversion symmetry, but in a higher dimension. One-dimensional systems have attracted much attention over the years. One of the most paradigmatic cases is the Su-Schrieffer-Heeger (SSH) model of a dimerized lattice, known as the simplest case of a topological insulator. , indicating that the system can host states of topological origin.

%Motivated by the ability of photonic lattices to realize various optical devices, a property which can be used for manipulating light

%\date{\today} 
\maketitle

\section{\label{Introduction}Introduction}
The understanding of topological states of matter~\cite{thouless_quantized_1982,Haldane_Nonlinear_1983,prange1987quantum,Wen_Topological_1995} has rapidly expanded over the last decade since the discovery of topological insulators~\cite{kane_quantum_2005,*kane_topological_2005,konig2007quantum,hasan_colloquium_2010,Qi_Topological_2011}. Examples range from topological states in driven quantum systems~\cite{Oka2009,lindner_floquet_2011,Foa_2014} to artificial systems including ultracold matter~\cite{atala_direct_2013,jotzu_experimental_2014} and photonic waveguides~\cite{hafezi_2011_robust,rechtsman_2013_photonic,ozawa_2018_topological}. Although originally most of the studies focused on two-dimensional (2D) systems~\cite{haldane_model_1988,kane_quantum_2005}, later on they evolved to three~\cite{hsieh_topological_2008} and one dimensions~\cite{Lang2012}.

Quasicrystals~\cite{Shechtman1984}, materials characterized by long-range orientational order but without the periodicity of crystals, appeared to be out of the topological chart until the work of Kraus and coworkers~\citep{Kraus2012}. They showed that lower dimensional quasiperiodic systems can feel the effect of a higher-dimensional “ancestor” crystal, through additional degrees of freedom $\phi$ that appear as remnants of the higher dimensionality. Shortly after, it was shown that crystal and quasicrystal band insulators are topologically equivalent~\cite{Madsen_2013}. Because of these connections, together with the reduced complexity of low-dimensional lattices and also thanks to the advance in experimental techniques, research on topological states in 1D systems has been reignited. Indeed, one can probe these states in ultracold atoms~\cite{atala_direct_2013,He2018}, photonic crystals~\cite{Kraus2012,Verbin_2013} and even in photonic Fibonacci quasicrystals~\cite{Verbin_2015}. In addition, adiabatic topological pumping~\cite{Kraus2012,Verbin_2015,ke2016topological} and discrete-time quantum walks~\cite{Kitagawa_2010,Wang_2017} have been investigated in 1D lattices. %Also, experimental signatures of Majorana fermions have been identified in several 1D topological superconductors in the past few years~\cite{Mourik2012signatures,Das2012zero,Nadj2014observation}.

In this context, the study of trimer lattices has been object of considerable theoretical and experimental interest because of their unique physical properties and very rich phase diagram~\cite{Macedo1995,Montenegro2014,Jin2017,liu2017new}. 
Particularly, it has been reported that the existence of edge states located only  at one edge/end of a Hermitian trimerized lattice, is due to a symmetry of the unit cell, which makes the Berry phase to be piecewise continuous rather than  discrete~\cite{liu2017new}. However, in a typical 1D Hermitian topological insulator, topological invariants takes only one value of a discrete set of values, and the bulk-boundary correspondence ensures the appearance of, at least, a pair of localized edge states: one on the left and one on the right. In fact, the capacity to exhibit a single edge state on one side of a system has been attributed to non-Hermitian systems~\cite{lee_anomalous_2016}, where the bulk-boundary correspondence is subject of intense debate and controversy~\cite{lee_anomalous_2016,Xiong_Why_2018,Martinez_PRB_2018,Ueda_2018,*[][{, and references therein. }]martinez2018topological}.

Motivated by the ability of photonic lattices to realize various optical devices~\cite{Garanovich_2012}, here we consider a simple and paradigmatic model consisting of a one-dimensional lattice with a three-site basis and examine in detail the edge states and their topological character. Interestingly, this model exhibits diverse in-gap edge states as the parameters are varied: (i) chiral edge states, i.e., states that are localized only on one edge of the system without a counterpart on the opposite edge at the same energy, and which can be used for manipulating fundamental properties of light  in a controllable way, (ii) usual topological states (one on each edge at the same energy).  Originally fulled by the first type of edge states, apparently violating the bulk-boundary correspondence we proceeded to find a subtle connection with the physics of superlattices. Henceforth, unless otherwise indicated, when talking about chiral edge states we actually mean the edge modes located only at a single edge and belonging to the inversion-symmetry broken phase of the trimer lattice.

In this context, one may wonder, is it possible to relate the existence of these chiral edge states with a topological invariant defined within the bulk? Indeed, as we will show later, the chiral edge states of the trimer chain can be related to bulk topological numbers, but defined in an effective two-dimensional parent system. Furthermore, we show that, the edge states in the inversion-symmetry broken phase of the trimer lattice turn out to be robust against disorder. We also find that the topological phase transition point of the trimer lattice correlates with the topological phase transition in its associated parent system, in which case the Chern numbers are duplicated and their signs change.

\begin{figure}
\centering
\subfloat{\label{Fig_1a}\includegraphics[width=1.0\linewidth]{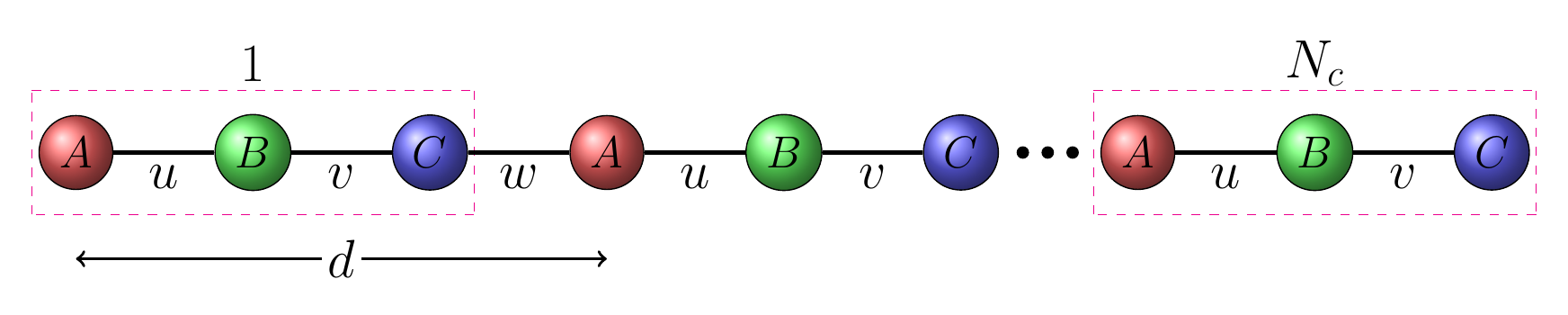}}
\caption{Scheme representing a finite section of the trimerized lattice model with $N_c$ unit cells, where $u$ and $v$ are the  intra-cell hopping amplitudes, $w$ is the inter-cell hopping amplitude and $d$ is the lattice spacing. Each unit cell contains three sites, $A$,  $B$ and $C$.}
\label{Fig_1}
\end{figure}  

The paper is organized as follows. In Section~\ref{Trimer}, we discuss general properties of the edge states of the trimer lattice, with special emphasis on those states that are localized only on one edge of the sample without a counterpart on the opposite edge. Moreover, in the Appendix we apply the recursive boundary Green function method to study the regions in the parameter space where edge states appear. In Section~\ref{Topological phase}, we analyze the symmetry exhibited by the trimer lattice and study its consequences for the edge states through the calculation of the Zak phase~\cite{Zak_1989}. In Section~\ref{Edge states}, we show that the edge states in the inversion-symmetry broken phase of the trimer lattice turn out to be robust to large amounts of disorder. In Section~\ref{Bulk-boundary}, we find a nontrivial correspondence between the chiral edge states of the trimer chain, with that of an effective 2D model (Aubry-Andr\'{e}-Harper model), which shows that the system can host states of topological origin, in much the same way as those in graphene ribbons with zigzag edges~\cite{Delplace_2011}. Also, we show that due to this subtle connection, the topological phase transition point of the trimer lattice allow us to find the corresponding topological phase transition point in its associated (two-dimensional) parent system. Finally, Section~\ref{Conclusions} provides a summary and concluding remarks.

%%%%%%%%%%%%%%%%%%%%%%%%%%%%%%%%%%%%%%%%%%%%%%%%%%%%%%%%%%%%%%%%%%%%%%%%%%%%%%%%%%%%%%%%%%%%%%%%%%%%%%%%%%%%%%%%%%%%%%%%%%%%%%%%%%%%%%%%%%%%%%%%%%%%%%%%%%%%%%%%%%%%%%%%%%%%%%%%%%%%%%%%%%%%%%%%%%%%%%%%%%%%%%%%%%%%%%%%%%%% Section II %%%%%%%%%%%%%%%%%%%%%%%%%%%%%%%%%%%%%%%%%%%%%%%%%%%%%%%%%%%%%%%%%%%%%%%%%%%%%%%%%%%%%%%%%%%%%%%%%%%%%%%%%%%%%%%%%%%%%%%%%%%%%%%%%%%%%%%%%%%%%%%%%%%%%%%%%%%%%%%%%%%%%%%%%%%%%%%%%%%%%%%%%%%%%%%%%%%%%%%%%%%%%%%%%%%%%%
\section{\label{Trimer}Trimer chain} 
Consider a system of spinless (or spin-polarized) electrons hopping on a one-dimensional chain composed of $N_c$ unit cells. Each unit cell hosts three distinct sites, which we denote as $A$,  $B$ and $C$ (see illustration in  Fig.~\ref{Fig_1}), hence we can expect to find three bands. The length of the unit cell is set to unity ($d=1$). We can model this  trimerized lattice by the following  tight-binding Hamiltonian:
\begin{equation}
H=\sum_{n=1}^{N_c}(uc_{A,n}^{\dagger}c_{B,n}+v c_{B,n}^{\dagger}c_{C,n}+wc_{C,n}^{\dagger}c_{A,n+1}+\mathrm{h.c.}),
\label{Hamiltonian}
\end{equation} 
where $c_{\alpha,n}^{\dagger}$ ($c_{\alpha,n}$) denotes the creation (annihilation) operator at site $\alpha$ (which can be either $A$, $B$ or $C$ type) of the $n$-th unit cell, $u$ and $v$ are the intra-cell hopping amplitudes, whereas $w$ is the inter-cell hopping amplitude. Assuming periodic boundary conditions (discrete translational invariance) along the length of the chain and performing Fourier transform of creation/annihilation operators: ${\psi_{n}=(1/\sqrt{N_c})\sum_{k}e^{ikn}\psi_{k}}$  with ${\psi{_n}=(c_{A,n},c_{B,n},c_{C,n})^{T}}$, we can write the Hamiltonian in reciprocal space as $H=\sum_{k}\psi_{k}^{\dagger}H(k)\psi_{k}$, where
\begin{align}
H(k) = \left(\begin{array}{ccc}
0 & u & we^{-ik}\\
u & 0 & v\\
we^{ik} & v & 0\label{HPBC}
\end{array}\right),
\end{align}
 
\begin{figure}
\centering
\subfloat{\label{Fig_2a}\includegraphics[width=1.0\linewidth]{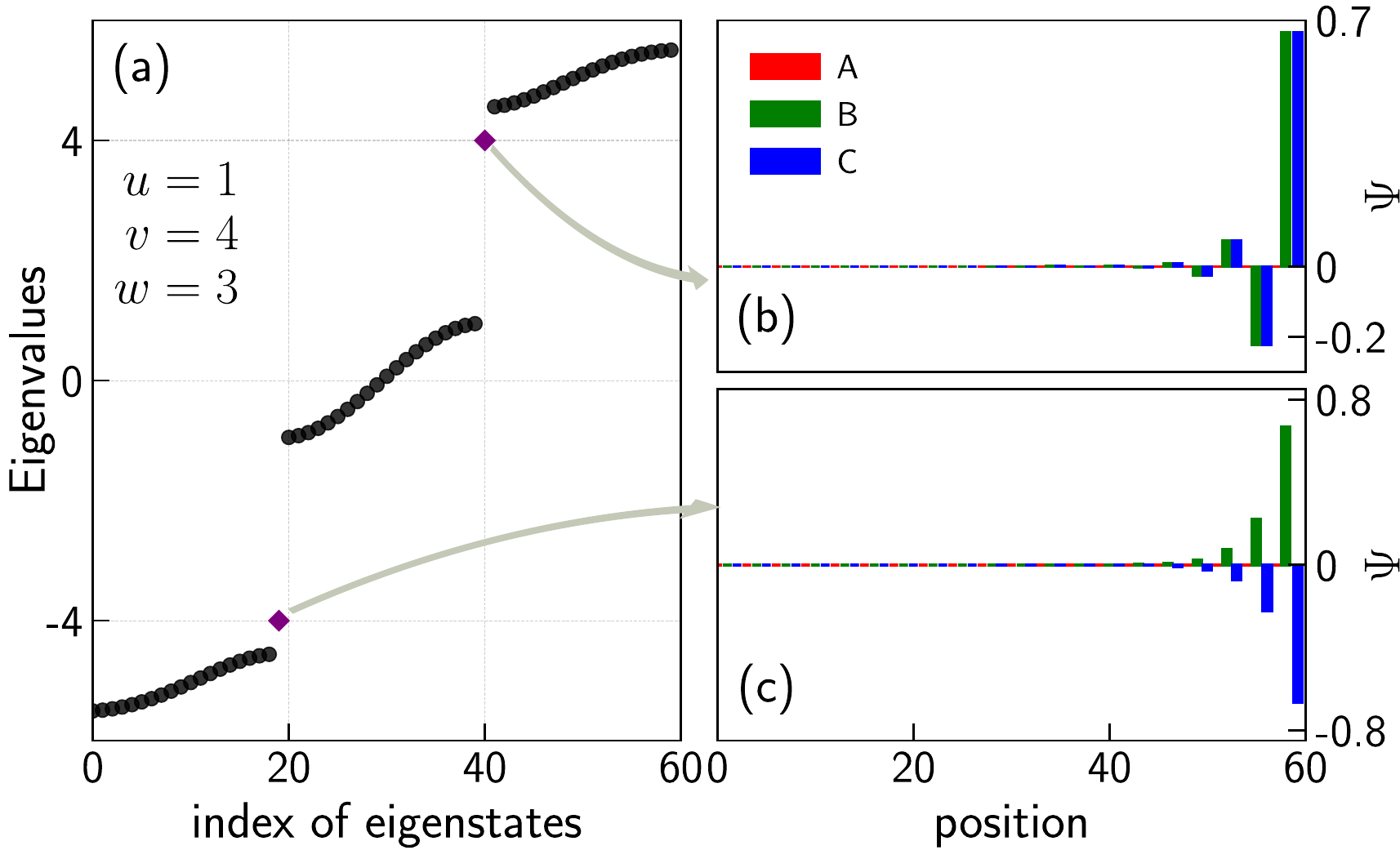}}
\subfloat{\label{Fig_2b}}
\subfloat{\label{Fig_2c}}
\caption{Energy spectrum and wavefunctions of the Hamiltonian of Eq.~(\ref{Hamiltonian}) with open boundary conditions for $N_c=20$  unit cells. (a) Energy spectrum of the system for intra-cell hopping amplitudes ${u=1}$, ${v=4}$  and inter-cell hopping amplitude ${w=3}$. (b) and (c) show the wavefunctions of the two edge states with energies $\varepsilon=\pm 4$, both marked as purple diamonds in (a), localized on the right boundary of the system, respectively.}
\label{Fig_2}
\end{figure}
The spectrum of the above Hamiltonian consists of three dispersive bands, which only touch each other at the boundaries of the first Brillouin zone (BZ), $k=0$ and $\pi$, when $|u|=|v|=|w|$, i.e., in absence of trimerization. As the parameters are shifted from that condition, two band gaps appear in the band structure, both with the same value. Interestingly, these band gaps may host very peculiar edge states as those shown in Fig.~\ref{Fig_2}, where we have plotted the energy spectrum of a finite trimer lattice for a set of hopping values of ${u=1}$, ${v=4}$ and ${w=3}$. Note the presence of two in-gap edge states, with energies ${\varepsilon=\pm 4}$, both localized on the right boundary of the system, with probability distribution at sites of type $B$ and $C$ [see Fig.~\ref{Fig_2}~\subref{Fig_2b} and \subref{Fig_2c}]. It is worth noting that our system does not exhibit chiral symmetry, and unlike the states in the Su-Schrieffer-Heeger (SSH) model~\cite{Su_1979,asboth_short_2016}, the states shown in  Fig.~\ref{Fig_2}~\subref{Fig_2b} and \subref{Fig_2c} are chiral in the sense that they are present on one edge of the sample, but not on the opposite edge. In this context, as we will show in Section~\ref{Bulk-boundary}, these states can be interpreted as inherited from a higher dimension, through a mapping onto an effective 2D model, which in fact presents robust chiral states along the edges.

\begin{figure}
\centering
\subfloat{\label{Fig_3a}\includegraphics[width=1.0\linewidth]{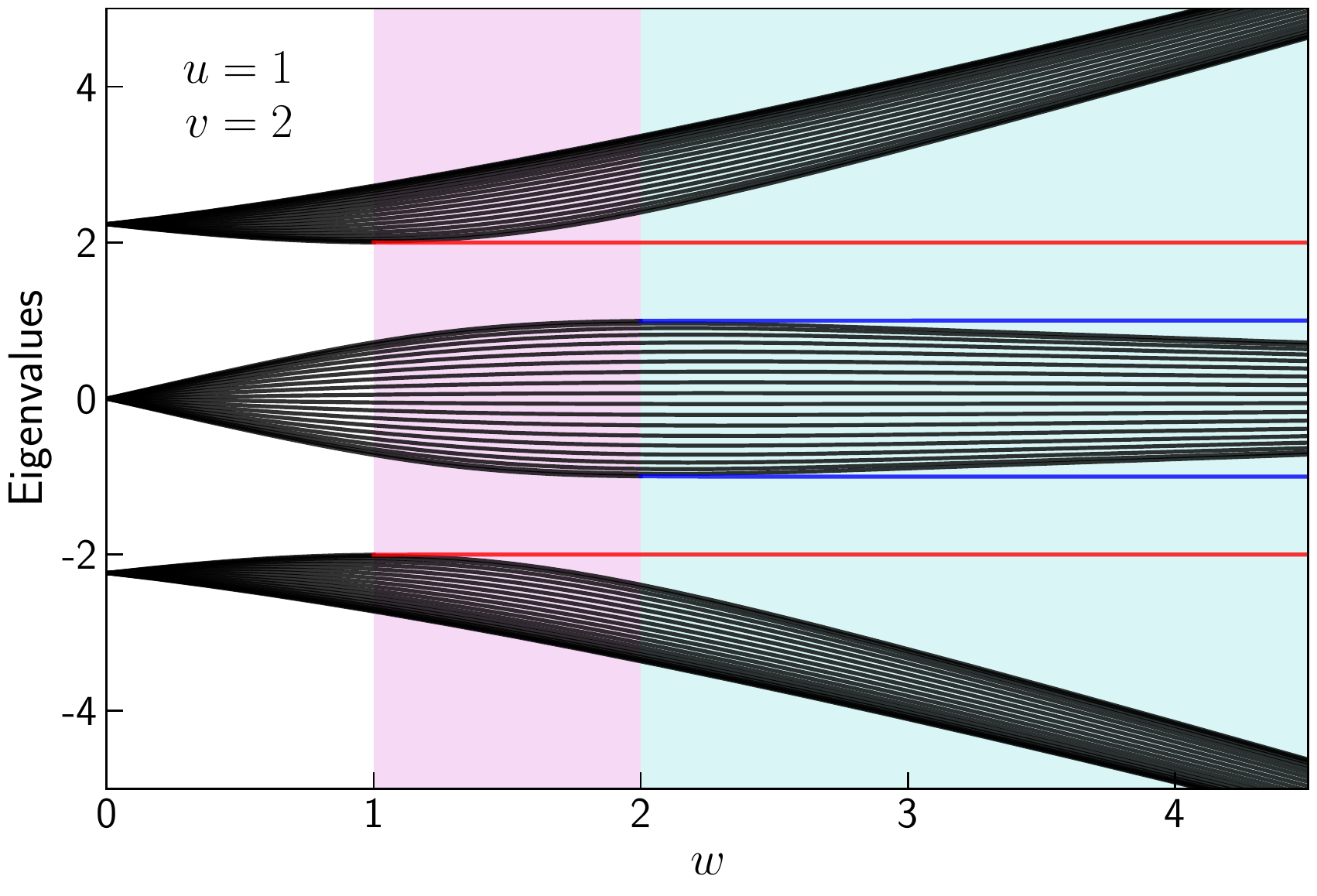}}
\subfloat{\label{Fig_3b}}
\caption{Energy spectrum of the system under open boundary condition as a function of the inter-cell hopping amplitude $w$, for $N_c=20$  unit cells and intra-cell hopping amplitudes of ${u=1}$ and ${v=2}$ (inversion-symmetry broken phase). We highlight three different regions with: (white) no edge states (${w<u,v}$), (light magenta) two in-gap edge states localized on the right boundary (${u<w<v}$), and  (light cyan)  two pairs of edge states localized on both edges of the system (${u,v<w}$).}
\label{Fig_3}
\end{figure}

In Fig.~\ref{Fig_3} we show the energy spectrum of a finite trimer lattice with $N_c=20$ unit cells, as a function of the inter-cell hopping amplitude $w$, and with intra-cell hopping amplitudes of ${u=1}$ and ${v=2}$. We can observe a different  number of in-gap edge states emerging  as $w$ changes. A closer analysis reveals three different regions, as shown in Fig.~\ref{Fig_3}. The first one (in white) with no edge states (${w<u,v}$), the second one (in light magenta) with two edge states localized on the right boundary of the system (${u<w<v}$), and the last one (in light cyan) with two pairs of edge states localized on both ends of the system (${u<v<w}$). Depending on whether ${u<w<v}$ or ${v<w<u}$, edge states will appear localized on the right or left boundaries of the system, respectively. Here we assume that ${0<u,v,w}$ for simplicity, but results can be easily derived in the general case. In fact, we use a recursive boundary Green function method~\cite{Peng2017}  (see the Appendix for details) to study the regions in the parameter space where edge states appear in a general framework.

This apparent violation of the bulk-boundary correspondence, also reported in Ref.~[\onlinecite{liu2017new}] as two new phases characterized by piecewise continuous Berry phases,  motivates us to look closer to this problem. In the following, we explore the localization properties and topological nature of the in-gap states.

%%%%%%%%%%%%%%%%%%%%%%%%%%%%%%%%%%%%%%%%%%%%%%%%%%%%%%%%%%%%%%%%%%%%%%%%%%%%%%%%%%%%%%%%%%%%%%%%%%%%%%%%%%%%%%%%%%%%%%%%%%%%%%%%%%%%%%%%%%%%%%%%%%%%%%%%%%%%%%%%%%%%%%%%%%%%%%%%%%%%%%%%%%%%%%%%%%%%%%%%%%%%%%%%%%%%%%%%%%%% Section III %%%%%%%%%%%%%%%%%%%%%%%%%%%%%%%%%%%%%%%%%%%%%%%%%%%%%%%%%%%%%%%%%%%%%%%%%%%%%%%%%%%%%%%%%%%%%%%%%%%%%%%%%%%%%%%%%%%%%%%%%%%%%%%%%%%%%%%%%%%%%%%%%%%%%%%%%%%%%%%%%%%%%%%%%%%%%%%%%%%%%%%%%%%%%%%%%%%%%%%%%%%%%%%%%%%%%%
\section{\label{Topological phase} Inversion-symmetric trimer chain} 

A general result of the classification of non-interacting fermionic topological phases~\cite{Schnyder_Classification_2008,Kitaev_periodic_2009,Ryu_topological_2010}, is that topological phases of matter in 1D can only exist through the imposition of symmetries on the system. Hence, in 1D and in the presence of either chiral~\cite{Schnyder_Classification_2008,Kitaev_periodic_2009,Ryu_topological_2010,Lieu_2018,Yuce_2018} or inversion symmetry~\cite{Hughes_Inversion_2011}, topological phases will be protected as long as the symmetry is preserved. Therefore, it is convenient to analyze the symmetries of the Hamiltonian, which will allow us to determine if there is any symmetry protected topological phase. 

Similarly to the chirally symmetric (Hermitian) SSH model, the topological properties of the trimerized chain are regulated by the relative strength of the inter-cell and intra-cell hopping amplitudes. In contrast, important distinct features appear since the unit cells of the two models are different. For $u=v$, $H(k)$ is inversion symmetric, with the inversion center lying at the midpoint $B$ between two sites $A$ and $C$ within a unit cell, i.e., ${\cal P}H(k){\cal P}^{-1}=H(-k)$, where the inversion operator
\begin{equation}
{\cal P}=\left(\begin{array}{ccc}
0 & 0 & 1\\
0 & 1 & 0\\
1 & 0 & 0
\end{array}\right),\,\,\,\,{\cal P}^{2}=1;\,\,\,\,{\cal P}={\cal P}^{-1},
\end{equation}
plays the equivalent role as the $\sigma_x$ operator for the SSH model~\cite{asboth_short_2016}.

As mentioned before, if the spatial dimension is one and if no symmetry is assumed, there are no topological phases~\cite{Schnyder_Classification_2008,Kitaev_periodic_2009,Ryu_topological_2010}, i.e., all gapped Hamiltonians in 1D are equivalent to the same trivial phase. However, in 1D and in the presence of inversion symmetry $\cal P$, one can classify different insulators by $\mathbb{Z}$, through a quantized topological index which can take only the values $0$ or $\pi$ (modulo $2\pi$), denoting the trivial and nontrivial topological insulators, respectively. This 1D topological invariant, which is intimately related to the existence of edge states through the bulk-boundary correspondence, is usually called the Zak phase~\cite{Zak_1989} and is defined as $\mathcal {Z}=i\int_{-\pi}^{\pi}dk{\braket{\psi_{k} | \partial{_k} \psi_{k}}}$, where $\psi_{k}$  are the Bloch wavefunctions. The Zak phase for the lower band can be computed and is found to be
\begin{equation}
  \mathcal{Z} =
  \begin{cases}
    0 & \text{if $|u|=|v|>|w|$} \\
    \pi & \text{if $|u|=|v|<|w|$}
  \end{cases}.
\end{equation}

\begin{figure}
\subfloat{\label{Fig_4a}\includegraphics[width=1.0\linewidth]{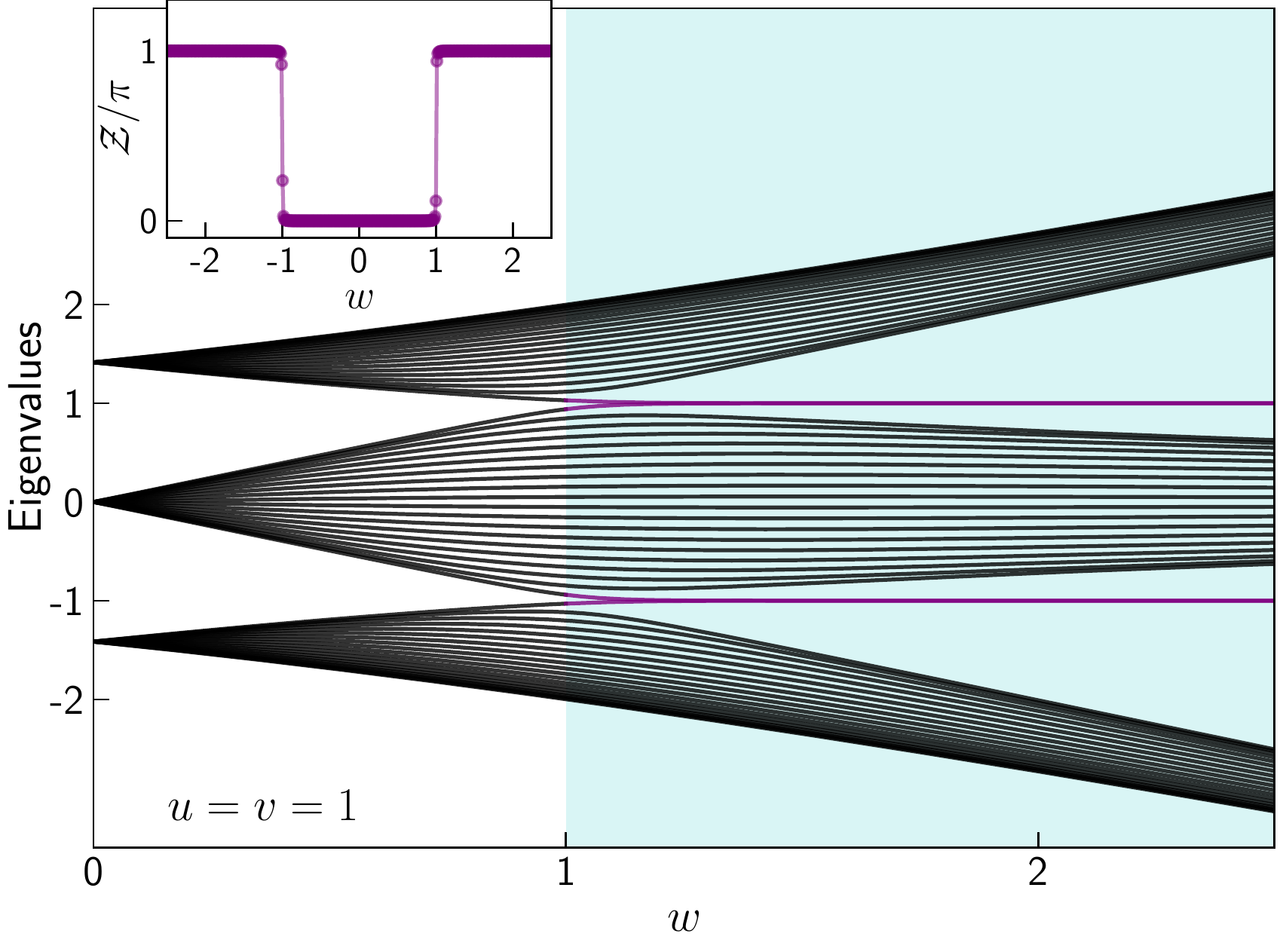}}
\caption{Energy spectrum and Zak phase (inset) of the system under open boundary condition as a function of the inter-cell hopping amplitude $w$,  for $N_c=20$  unit cells and intra-cell hopping amplitudes ${u=v=1}$ (inversion-symmetric phase). Note that, when the inversion symmetry is preserved, $|w|>1$ ($|w|<1$) corresponds to the topological (non-topological) phases of the trimer lattice.}
\label{Fig_4}
\end{figure}

A nontrivial Zak phase implies that a pair of topologically protected edge states will appear at the boundaries of the system, one on the right and one on the left when we cut the chain. Fig.~\ref{Fig_4} shows the spectrum of the Hamiltonian under open boundary conditions for ${|u|=|v|=1}$ (inversion-symmetric phase) as a function of $w$. A direct manifestation of the nontrivial Zak phase of $\pi$ are the degenerate gapless modes, with energies ${\varepsilon=\pm 1}$, appearing at the boundaries of the system. As we can see the two regions ${|u|=|v|<|w|}$ and vice versa correspond to two topologically distinct phases, i.e., the system undergoes a topological phase transition with gap closing at ${|u|=|v|=|w|}$, from a trivial insulator to an inversion symmetry protected topological insulator. This is a common example of a topological phase transition with gap closing, as one cannot continuously switch between the two phases without either closing the bulk energy gaps or breaking the inversion symmetry. It is worth noting that, the spectrum depicted in Fig.~\ref{Fig_4} resembles that of the SSH model very closely. In fact, both are symmetrically arranged around zero energy and have in-gap edge states protected by inversion or chiral symmetry, which appear after the gap-closing-and-reopening transition~\cite{asboth_short_2016}. 
%%%%%%%%%%%%%%%%%%%%%%%%%%%%%%%%%%%%%%%%%%%%%%%%%%%%%%%%%%%%%%%%%%%%%%%%%%%%%%%%%%%%%%%%%%%%%%%%%%%%%%%%%%%%%%%%%%%%%%%%%%%%%%%%%%%%%%%%%%%%%%%%%%%%%%%%%%%%%%%%%%%%%%%%%%%%%%%%%%%%%%%%%%%%%%%%%%%%%%%%%%%%%%%%%%%%%%%%%%%% Section IV %%%%%%%%%%%%%%%%%%%%%%%%%%%%%%%%%%%%%%%%%%%%%%%%%%%%%%%%%%%%%%%%%%%%%%%%%%%%%%%%%%%%%%%%%%%%%%%%%%%%%%%%%%%%%%%%%%%%%%%%%%%%%%%%%%%%%%%%%%%%%%%%%%%%%%%%%%%%%%%%%%%%%%%%%%%%%%%%%%%%%%%%%%%%%%%%%%%%%%%%%%%%%%%%%%%%%%
\section{\label{Edge states} Edge States and robustness to disorder} 

Because of the bulk-boundary correspondence in Hermitian systems, nontrivial topological invariants imply the existence of gapless states exponentially localized at the boundaries of the sample (gap closing at the transition, see Fig.~\ref{Fig_4}). Thus, when the inversion symmetry is broken, it seems natural to expect the disappearance or at least the instability of these states. However, as seen in  Figs.~\ref{Fig_2} and \ref{Fig_3}, there are still in-gap states which are localized only at one boundary of the system without an equivalent one at the opposite end. Also, as the robustness against disorder is a characteristic feature of edge states in the topological insulator phase, a natural question arises as to whether these chiral edge states are robust to disorder or not, since they are not protected by any symmetry. 

\begin{figure}
\centering
\subfloat{\label{Fig_5a}\includegraphics[width=1.0\linewidth]{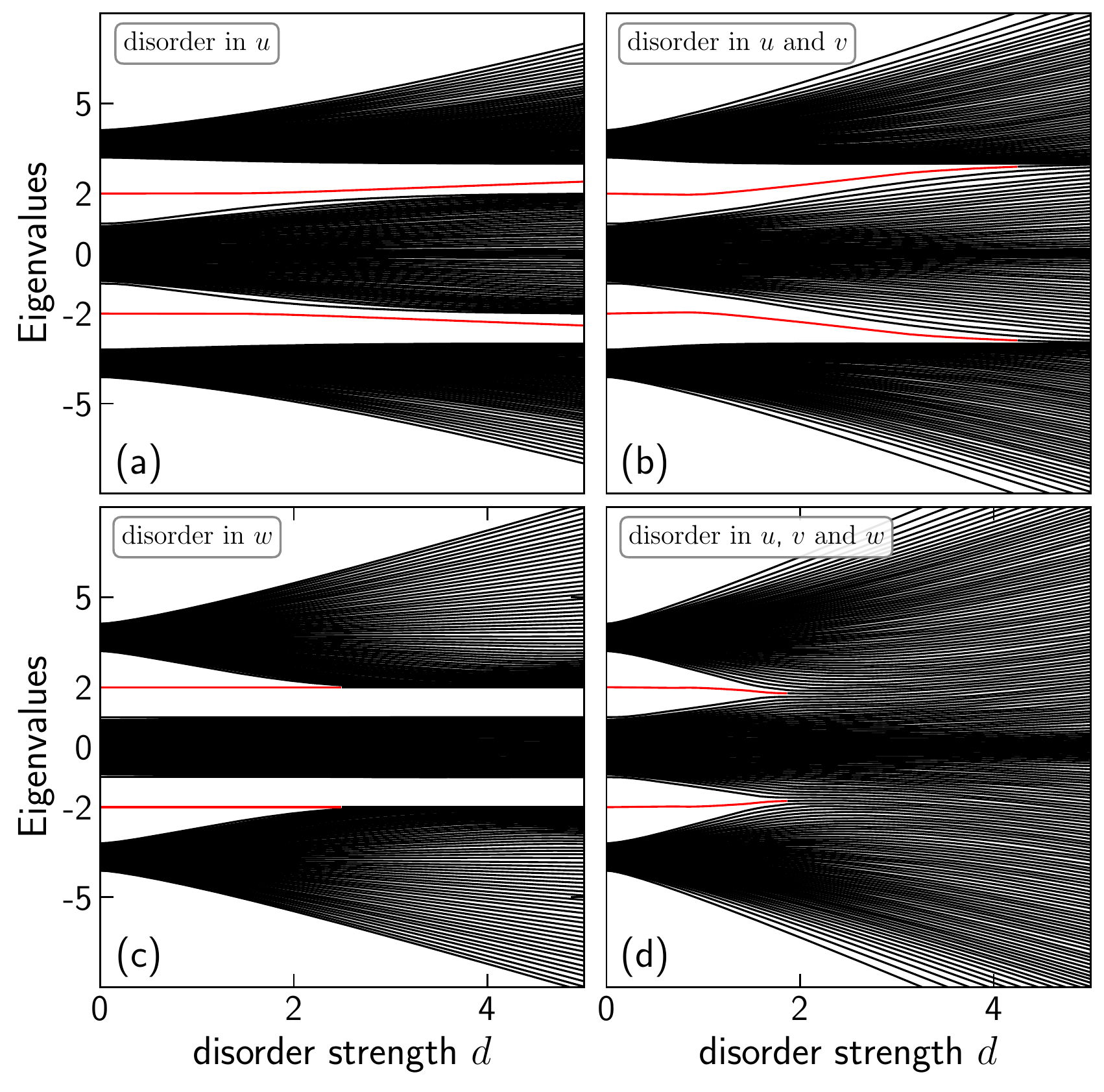}}
\subfloat{\label{Fig_5b}}
\subfloat{\label{Fig_5c}}
\subfloat{\label{Fig_5d}}
\caption{Energy spectrum of a finite trimer lattice with $N_c=60$ unit cells, under the effect of increasing amounts of disorder $d$: (a) in $u$, (b) in $u$ and $v$, (c) in $w$, and (d) in $u$, $v$ and $w$. The starting points for all plots are $u=1$, $v=2$ and $w=3$ (inversion-symmetry broken phase), and the results are the average over 100 simulations.}
\label{Fig_5}
\end{figure}

In order to gain insight into this question, we numerically investigate the robustness of the edge states against disorder in the inversion-symmetry broken phase of the trimer lattice. The Hamiltonian of Eq.~(\ref{Hamiltonian}) contains three hopping parameters: $u$, $v$ and $w$ on which we introduce random disorder, i.e., $u_n=u+d\gamma_n$, $v_n=v+d\gamma_n$ and $w_n=w+d\gamma_n$,   where $n$ is the cell index, $\gamma_n$ is uniformly distributed between  $1$ and $-1$ and $d$ is the disorder strength. For the sake of generality, we also allow the disorder to be different for each unit cell, and act on the hopping parameters: $u$, $v$ and $w$, independently.    

Fig.~\ref{Fig_5} shows the results, averaged over 100 simulations, for four types of disorder as a function of the disorder strength $d$.  The starting points for all plots are $u=1$, $v=2$ and $w=3$. Disorder in the intra-cell hopping $u$ and inter-cell hopping $w$, are depicted in Fig.~\ref{Fig_5}~\subref{Fig_5a} and \ref{Fig_5}~\subref{Fig_5c}, respectively. On the other hand, in Fig.~\ref{Fig_5}~\subref{Fig_5b} we allow disorder to act on $u$ and $v$, independently. Lastly, Fig.~\ref{Fig_5}~\subref{Fig_5d} shows the spectrum of the system with disorder acting in $u$, $v$ and $w$, also independently.  

We can observe that, even when details of the spectrum are modified, and the robustness of the edge states against the disorders are slightly different, the localized nature of the  edge states remain. Moreover, the eigenvalues (of the right edge states) remain at $\varepsilon=\pm v$ until the disorder is strong enough to either, cause the eigenvalues reach the band (becoming extended) or close the band gap. Remarkably, these edge states are robust to large amounts of disorder and, as we will see next, we can associate a topological origin to them, originated from a map onto an effective model in a higher dimension. This contrasts with the case of the edge states of zigzag graphene ribbons, which have a topological origin rooted in a lower dimension~\cite{Delplace_2011}, but are fragile to disorder.

%%%%%%%%%%%%%%%%%%%%%%%%%%%%%%%%%%%%%%%%%%%%%%%%%%%%%%%%%%%%%%%%%%%%%%%%%%%%%%%%%%%%%%%%%%%%%%%%%%%%%%%%%%%%%%%%%%%%%%%%%%%%%%%%%%%%%%%%%%%%%%%%%%%%%%%%%%%%%%%%%%%%%%%%%%%%%%%%%%%%%%%%%%%%%%%%%%%%%%%%%%%%%%%%%%%%%%%%%%%% Section V %%%%%%%%%%%%%%%%%%%%%%%%%%%%%%%%%%%%%%%%%%%%%%%%%%%%%%%%%%%%%%%%%%%%%%%%%%%%%%%%%%%%%%%%%%%%%%%%%%%%%%%%%%%%%%%%%%%%%%%%%%%%%%%%%%%%%%%%%%%%%%%%%%%%%%%%%%%%%%%%%%%%%%%%%%%%%%%%%%%%%%%%%%%%%%%%%%%%%%%%%%%%%%%%%%%%%%
\section{\label{Bulk-boundary} Topological Origin of edge states in the inversion-symmetry broken phase}

The bulk-boundary correspondence dictates the existence of gapless edge states from bulk topological invariants. Remarkably, depending on the values of the hopping amplitudes, the trimer lattice can exhibit diverse in-gap edge states which \quotes{apparently} violate  the bulk-boundary correspondence, i.e., states appear localized only at one end of the system without a counterpart at the opposite end. This, combined with the fact that these edge states turn out to be robust against disorder, makes one  wonder about the bulk-boundary correspondence, and the nature/origin of these edge states in such a system.

To answer these questions and motivated by the fact that families of 1D band insulators, i.e., systems where the Hamiltonian depend periodically on a parameter $\phi$ (defining an effective 2D model), share the same topological classification as the quantum Hall effect~\cite{Thouless_1983,Kraus2012,Madsen_2013}, we model the trimer lattice through the so-called commensurate off-diagonal Aubry-Andr\'{e}-Harper (AAH) model~\cite{harper1955single,aubry1980analyticity}. This model can be described by the following tight-binding Hamiltonian
\begin{equation}
H=\sum_{n=1}^{N}t[1+\lambda\cos(2\pi bn+\phi)c_{n+1}^{\dagger}c_{n}]+\mathrm{h.c.},\label{AAH}
\end{equation}
where $N$ is the number of lattice sites, $c_{n}^{\dagger}$ ($c_{n}$) is the creation (annihilation) operator at site $n$, $t$ is the hopping amplitude which is set to be the unit of the energy ($t=1$), and the parameter $\lambda$ is the modulation amplitude of the coupling strength. The modulation periodicity is controlled by $b=p/q$ ($p$ and $q$ are coprime numbers),  leading to a commensurate (incommensurate) modulation  with the lattice, whenever $b$ is rational (irrational). Here, our interest is to discuss the case of $b$ rational, or more specifically, $p=1$ and $q=3$ (${b\equiv1/3}$) which leads to a trimerized model with three bands. On the other hand, the phase factor $\phi$ plays the role of an additional degree of freedom, which in our case, allow us to obtain a whole family of different trimers. 

A family of trimers, i.e., $\left\{ H(\phi)|0\leq\phi<2\pi\right\}$, defines an effective model in two dimensions. Therefore, identifying $\phi$  as one component of the wave vector of a 2D system, we can map this 1D model to a 2D model, in such a way that the topological properties of our 1D trimer chain can be easily studied by using topological concepts for 2D systems~\cite{Lang2012,Kraus2012}. Hence, assuming periodic boundary conditions on the system, Eq.~(\ref{AAH}),  and performing Fourier transforms of creation/annihilation operators, Chern numbers for individual ($n$-th) bands can be defined in an effective 2D space, ($k,\phi$), over the BZ $(\ensuremath{0\leq k<2\pi/q},\,0\leq\phi<2\pi)$  as
\begin{equation}
\nu_{n}=\frac{1}{2\pi}\int_{0}^{2\pi/q}dk\int_{0}^{2\pi}d\phi\left(\partial_{k}A_{\phi}-\partial_{\phi}A_{k}\right),
\end{equation}
with the Berry connection $A_{r}=i$ ${\braket{\psi(k,\phi)|\partial{_{r}}|\psi(k,\phi)}}$ ($r=k,\phi$), where $\psi(k,\phi)$ are the Bloch wavefunctions. We numerically calculate the Chern numbers for individual bands as a function of $\lambda$. We found that, when the modulation amplitude, $\lambda<\lambda_c$, the Chern numbers are $(\nu_{1},\nu_{2},\nu_{3})=(-1,2,-1)$, while for $\lambda>\lambda_c$ the Chern numbers are $(\nu_{1},\nu_{2},\nu_{3})=(2,-4,2$), with $\lambda_c=4$, predicting a topological phase transition where not only are the Chern numbers doubled, but they also change in sign, in agreement with results derived in the context of Thouless pumping~\cite{ke2016topological}. As noted in Ref.~\cite{ke2016topological} this means that, if we perform a particle pumping experiment along the lattice, the propagation direction will change by the opposite, and it will be faster.

\begin{figure}
\centering
\subfloat{\label{Fig_6a}\includegraphics[width=1.0\linewidth]{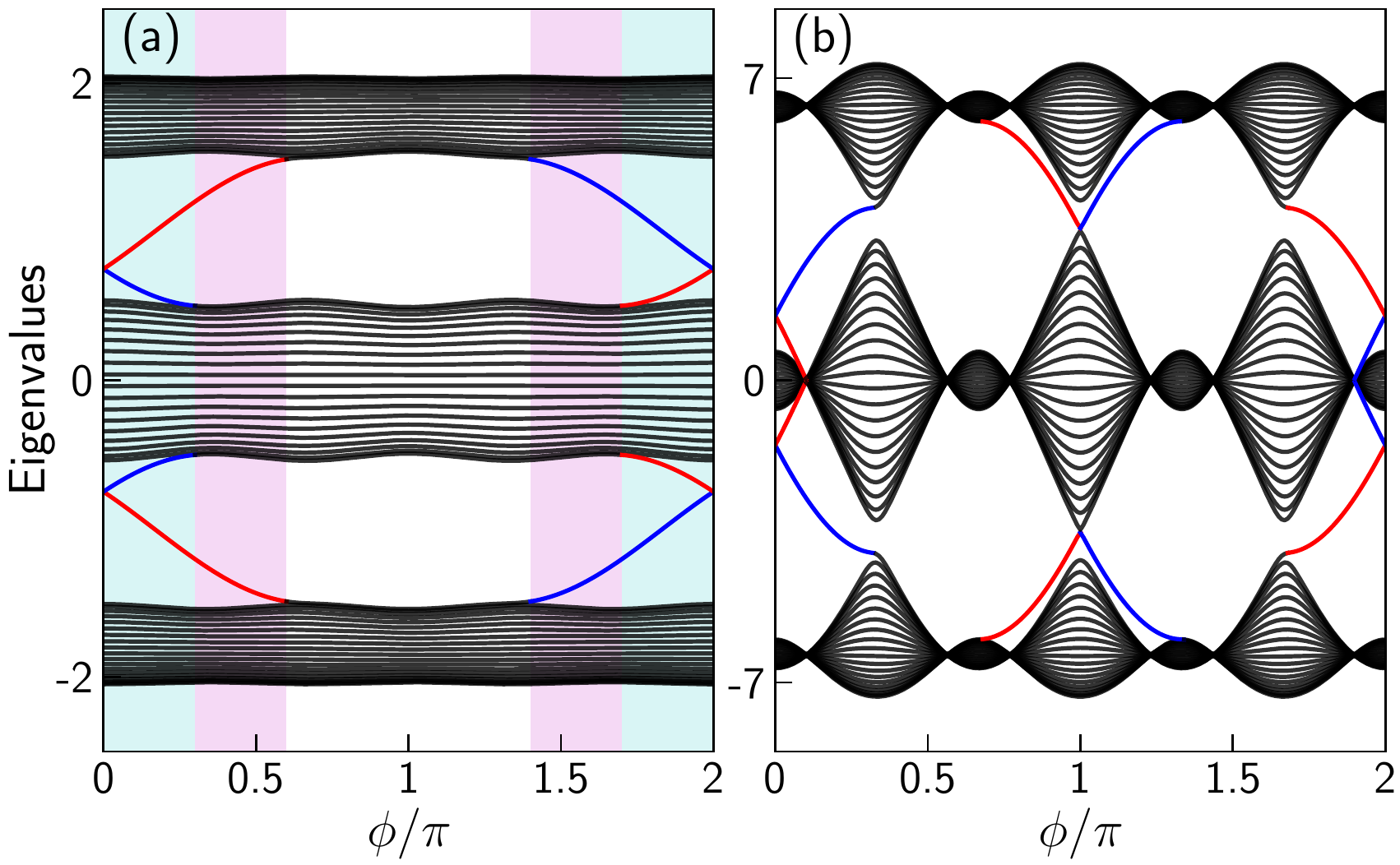}}
\subfloat{\label{Fig_6b}}
\caption{Energy spectrum  of the commensurate off-diagonal AAH model, Eq.~(\ref{AAH}), under open boundary condition as a function of $\phi$, for $b=1/3$, $N=60$ sites, and two different values of $\lambda$: (a) $\lambda=0.5$ and (b) $\lambda=5$. The bulk Bloch states correspond to black lines, whereas the edge states localized on the right (left) boundary correspond to red (blue) lines. Note the correspondence between the highlighted areas in (a) and those in Fig.~\ref{Fig_3}. Chern numbers for individual bands are: (a) $(\nu_{1},\nu_{2},\nu_{3})=(-1,2,-1)$ before the transition and (b) $(\nu_{1},\nu_{2},\nu_{3})=(2,-4,2$) after the transition point, $\lambda_c=4$.}
\label{Fig_6}
\end{figure}

Fig.~\ref{Fig_6} shows the energy spectrum of the commensurate off-diagonal AAH model under open boundary condition for a lattice of finite length, $N=60$, and two values of $\lambda$: Fig.~\ref{Fig_6}~\subref{Fig_6a} for $\lambda=0.5$ and Fig.~\ref{Fig_6}~\subref{Fig_6b} for $\lambda=5$. The bulk Bloch states are denoted with black lines whereas the edge states localized at the right (left) boundary correspond to red (blue) lines. By keeping $b=1/3$ and scanning $\phi$ from $0$ to $2\pi$ one can observe in Fig.~\ref{Fig_6}~\subref{Fig_6a} that the two gaps are closed by a few modes, reproducing all possible configurations that appear in Fig.~\ref{Fig_3}. Indeed, each value of $\phi\in[0,2\pi)$ corresponds to a certain set of hopping values, that is, each cut as a function of $\phi$ defines a trimer. In Fig.~\ref{Fig_6}~\subref{Fig_6a} we have highlighted in white, light magenta and light cyan the corresponding regions where there are no edge states, two edge states located at the right (red line) or left (blue line) boundaries of the system, and two pairs of edge states localized on both ends of the system, respectively. It is worth noting the correspondence between the highlighted areas in Fig.~\ref{Fig_6}~\subref{Fig_6a} and those in Fig.~\ref{Fig_3}, in terms of localization and number of edge states. Also notice that the number edge modes in Fig.~\ref{Fig_6}~\subref{Fig_6b} is twice of that in Fig.~\ref{Fig_6}~\subref{Fig_6a}, while the propagation direction is the opposite.

The individual members of the trimer's family are not topological in general, although they exhibit states located at the boundaries of the system. As discuses in Section~\ref{Topological phase}, only when the inversion symmetry is preserved, an element of the family is topological. These are the cases that have a crossing at the inversion-symmetry point, indicating the emergence of degenerate edge states protected by the inversion symmetry. On the other hand, only when scanning $\phi$ from zero to $2\pi$ we can obtain all the possible variants that the trimer lattice can assume. This suggest that the appearance of the chiral states has a topological origin because they are associated with a whole family, and only as a whole family we can define invariants (Chern numbers) and establish a bulk-boundary correspondence.

As mentioned before, as the hopping amplitude $\lambda$ increases, the system undergoes a topological phase transition~\cite{ke2016topological}. The topological phase transition point can be found, analytically, by solving a cubic equation that results from Eq.~(\ref{AAH}) assuming periodic boundary conditions and $b=1/3$~\cite{ke2016topological}. Here we put forward an interesting connection between the transition point in 1D and 2D, whereby the precise determination of the closing gap and the transition point in 1D, correctly determines the corresponding position of closing gap and  transition point in 2D, a property which can be extended to other systems with rational $b$ and odd $q$. 

The topological phase transition in the trimer lattice, as we saw in Section.~\ref{Topological phase}, occurs when $|u|=|v|=|w|$, thus, the position corresponding to the closing gap in the effective 2D system is not changed, only modulated. In this way, we can identify $u=1+\lambda\cos(2\pi /3+\phi)$, $v=1+\lambda\cos(4\pi /3+\phi)$ and  $w=1+\lambda\cos(\phi)$. The hopping amplitudes $u$ and $v$ are always equal when the phase factor is  $\phi=\pi$, then we have  $u=v=1+\lambda/2$ and $w=1-\lambda$. In fact, by equating 
\begin{equation}
|1+\lambda/2|=|1-\lambda|,
\end{equation}
we obtain the critical value $\lambda_{c}=4$ for the topological phase transition, in full agreement with that found in Ref.~\cite{ke2016topological} in the context of Thouless pumping of light in photonic waveguide arrays. Lastly, we want to emphasize that this connection reinforces our suggestion that the trimer lattice can host states of topological origin. 

%%%%%%%%%%%%%%%%%%%%%%%%%%%%%%%%%%%%%%%%%%%%%%%%%%%%%%%%%%%%%%%%%%%%%%%%%%%%%%%%%%%%%%%%%%%%%%%%%%%%%%%%%%%%%%%%%%%%%%%%%%%%%%%%%%%%%%%%%%%%%%%%%%%%%%%%%%%%%%%%%%%%%%%%%%%%%%%%%%%%%%%%%%%%%%%%%%%%%%%%%%%%%%%%%%%%%%%%%%%% Section VI %%%%%%%%%%%%%%%%%%%%%%%%%%%%%%%%%%%%%%%%%%%%%%%%%%%%%%%%%%%%%%%%%%%%%%%%%%%%%%%%%%%%%%%%%%%%%%%%%%%%%%%%%%%%%%%%%%%%%%%%%%%%%%%%%%%%%%%%%%%%%%%%%%%%%%%%%%%%%%%%%%%%%%%%%%%%%%%%%%%%%%%%%%%%%%%%%%%%%%%%%%%%%%%%%%%%%%
\section{\label{Conclusions}Conclusions}
Motivated by analogy with photonic lattices, we have studied the edge states of a one-dimensional trimer lattice and examined its characteristics in the phases with and without inversion symmetry protection. Remarkably, we have shown that the edge states in the inversion-symmetry broken phase of the trimer model may appear located at a \textit{single} edge. In particular, the emergence of degenerate edge states in the inversion-symmetric phase of the trimer model has been characterized through the calculation of the Zak phase. If the inversion symmetry is broken, we have demonstrated that the chiral edge states remain robust to large amounts of disorder. This contrasts, for example, with the case of the edge states of zigzag graphene ribbons which are less robust, and have a topological origin rooted in a lower dimension~\cite{Delplace_2011}.  In addition, through the mapping onto the commensurate off-diagonal Aubry-Andr\'{e}-Harper model, we have captured the essentials of the whole family of trimers, which allows us to establish a direct connection between chiral edge modes in the two models, including the calculation of Chern numbers in this effective two-dimensional model. We thus suggest that the chiral edge modes of the trimer lattice have a topological origin inherited from this effective mapping. We have established a nontrivial connection between the topological phase transition point in the trimer lattice to that in its associated effective two-dimensional parent system. The topological phase transition point found here is in full agreement with results derived in the context of topological Thouless pumping in photonic lattices~\cite{ke2016topological}. This nontrivial connection strengthens our suggestion that the trimer lattice, in the inversion-symmetry broken phase, can host states of topological origin. 

\begin{acknowledgments}
V. M. Martinez Alvarez is grateful to L.E.F. Foa Torres for interesting discussions. This work was supported by CNPq and FACEPE through the PRONEX program, and CAPES (Brazilian agencies).
\end{acknowledgments}

\appendix*
\section{\label{Recursive}Recursive boundary Green function}
In order to investigate the localization properties of the edge states and its relationship with the relative strength of the inter-and intra-cell hopping amplitudes of the trimer lattice Hamiltonian, Eq.~(\ref{Hamiltonian}), in this Appendix we apply the recursive boundary Green function method which relates the presence or absence of edge states to the fixed points of the recursion~\cite{Peng2017}. 

%eliminar  Hamiltonian, Eq.~(\ref{Hamiltonian})

Following the scheme laid out in Ref.~\cite{Peng2017}, we can extend the system by adding sites until we obtain a trimer with $N_c$ unit cells. This can be achieved in two ways, either by extending the system to the right or to the left. In any case, considering that the chain is long enough (large $N_c$-limit), we expect that the boundary Green function becomes independent of the number of unit cells, but not on the specific boundary. Indeed, as we show next, in the inversion-symmetry broken phase of the trimer lattice, the right and left boundary Green functions are different. 

The boundary Green function $G_{3N_c}$ of a trimer chain with $3N_c$ sites can be related to the boundary Green function $G_{3N_c-1}$ of a chain with $3N_c -1$ sites through the Dyson equation:
\begin{equation}
(g_{3N_c}^{-1}-V_{3N_c-1}G_{3N_c-1}V_{3N_c-1}^{\dagger})G_{3N_c}=\mathbb{I},\label{dyson}
\end{equation}
where $g_{3N_c}^{-1}=\mathbb{I}\varepsilon$ is the bare Green function, and $V_{3N_c-1}$ is associated with the hopping terms $u$, $v$ and $w$. In this way, by iterating (three times) the recursion Eq.~(\ref{dyson}) from the right boundary we obtain a recursion for trimer chains with a number of sites multiple of three: 
\begin{equation}
G_{3N_{c}}^R=\left(\varepsilon-v^{2}(\varepsilon-u^{2}(\varepsilon-w^{2}G_{3N_c-3}^R)^{-1})^{-1}\right)^{-1}.\label{Gright}
\end{equation}
Analogously, we can study the appearance of edges states localized on the left boundary of the trimer lattice by iterating (also three times) the recursion Eq.~(\ref{dyson}) from the left boundary
\begin{equation}
G_{3N_{c}}^{L}=\left(\varepsilon-u^{2}(\varepsilon-v^{2}(\varepsilon-w^{2}G_{3N_{c}-3}^{L})^{-1})^{-1}\right)^{-1}.
\end{equation}
Notice that the above equation can be obtained from Eq.~(\ref{Gright}) through the exchange $u\leftrightarrow v$.  In what follows, we will describe how to obtain the right boundary Green function ($G^R$) and then extend the results to the left boundary Green function ($G^L$) through the exchange $u\leftrightarrow v$.

It is convenient to rewrite the recursion Eq.~(\ref{Gright}) as $G_{3N_c}^R-G_{3N_c-3}^R=\beta(G_{3N_c-3}^R)$, where the $\beta$ function is given by 
\begin{equation}
\beta(x)=\left(\varepsilon-v^{2}(\varepsilon-u^{2}(\varepsilon-w^{2}x)^{-1})^{-1}\right)^{-1}-x.
\end{equation}
The zeros of the $\beta$ function define the fixed-point boundary Green function. One can write the solution for the above quadratic equation as follows
\begin{align}
x & =\frac{\varepsilon(w^{2}-u^{2}+\varepsilon^{2}-v^{2})}{2w^{2}(\varepsilon^{2}-v^{2})}\nonumber \\
 & \pm\sqrt{\left(\frac{\varepsilon(w^{2}-u^{2}+\varepsilon^{2}-v^{2}))}{2w^{2}(\varepsilon^{2}-v^{2})}\right)^{2}-\frac{(\varepsilon^{2}-u^{2})}{w^{2}(\varepsilon^{2}-v^{2})}},
\end{align}
and making an expansion in power series of $(\varepsilon\pm v)$ we obtain \begin{align}
 & G_{\mathrm{regular}}^{R}=\pm\frac{u^{2}-4v{}^{2}-w^{2}}{8vw^{2}}+\mathcal{O}(\varepsilon\pm v),\\
 & G_{\mathrm{singular}}^{R}=\frac{(w^{2}-u^{2})}{2w^{2}(\varepsilon\pm v)}\pm\frac{u^{2}-4v{}^{2}-w^{2}}{8vw^{2}}+\mathcal{O}(\varepsilon\pm v).
\end{align}
The above regular and singular (right) boundary Green functions characterize the absence and presence of edge states localized on the right boundary of the system, respectively. In fact, the two poles $\varepsilon=\pm v$ of the singular boundary Green function indicate the energies of the edge states.

We know that a fixed point is stable when $\beta'(x)<0$. Then, the region of the parameter space where edge states appear can be analyzed by studying the stability of the fixed-point Green functions under the recursion, which leads to
\begin{align}
\left.\beta'(G_{\mathrm{regular}}^{R})\right|_{\varepsilon=\pm v} & =\frac{w^{2}}{u^{2}}-1,\\
\left.\beta'(G_{\mathrm{singular}}^{R})\right|_{\varepsilon=\pm v} & =\frac{u^{2}}{w^{2}}-1.
\end{align}
The emergence of edges states located at the right (left) boundary of the trimer lattice, with energies ${\varepsilon=\pm v}$ (${\varepsilon=\pm u}$), are characterized by the singular boundary Green function $G_{\mathrm{singular}}^{R}$ ($G_{\mathrm{singular}}^{L}$), which is stable for $|u|<|w|$ ($|v|<|w|$). On the other hand, the regular boundary Green function $G_{\mathrm{regular}}^{R}$ ($G_{\mathrm{regular}}^{L}$), is stable for $|w|<|u|$ ($|w|<|v|$), which means that there are no edge states located at the right (left) boundary of the trimer lattice in this region of parameter space. 

We conclude this Appendix by mentioning that the above results fully support the findings reported in Section~\ref{Trimer}. 

\bibliographystyle{apsrev4-1_title}
\bibliography{bib.bib}
\end{document}